\documentclass[twocolumn,epjc3]{svjour3}          

\smartqed  

 \RequirePackage[T1]{fontenc}
 \RequirePackage{mathptmx}      
 \RequirePackage{flushend}
 \RequirePackage[numbers,sort&compress]{natbib}
 \RequirePackage[colorlinks,citecolor=blue,urlcolor=blue,linkcolor=blue]{hyperref}

 \usepackage{mathrsfs}
 \usepackage[mathscr]{euscript}
 \usepackage{amsmath}
 \usepackage{amssymb}
 \usepackage{upgreek}
 \usepackage[english]{babel} 
 \usepackage{csquotes} 

\renewcommand{\mathcal}{\mathscr}

\newcommand{\bnabla}{\boldsymbol{\nabla}}

\allowdisplaybreaks 

\begin{document}

\title
      {Extended Grimus--Stockinger theorem and inverse square law violation in quantum field theory}


\author{Vadim~A.~Naumov\thanksref{e1,addr1}
        \and
        Dmitry~S.~Shkirmanov\thanksref{e2,addr1} 
       }

\thankstext{e1}{e-mail: vnaumov@theor.jinr.ru}
\thankstext{e2}{e-mail: shkrimanov@theor.jinr.ru}

\institute{{Bogoliubov Laboratory of Theoretical Physics, Joint Institute for Nuclear Research, RU-141980 Dubna, Russia}\label{addr1}}

\date{Received: date / Accepted: date}

\maketitle

\begin{abstract}

We study corrections to the Grimus-Stockinger theorem dealing with the large-distance asymptotic behavior of the
external wave-packet modified neutrino propagator within the framework of a field-theoretical description of
the neutrino oscillation phenomenon. The possibility is discussed that these corrections, responsible for breakdown
of the classical inverse-square law (ISL), can lead to measurable effects at small but macroscopic distances
accessible in the SBL (anti)neutrino experiments and in particular can provide an explanation of the well-known
reactor antineutrino anomaly.

\end{abstract}

\section{Introduction}
\label{sect:Introduction}

The neutrino oscillation phenomenon (transmutation of one neutrino flavor to another in passing from a source to a target)
is regarded as the most likely explanation of the results of the last decades experiments with solar, atmospheric, reactor,
and accelerator neutrinos and antineutrinos.
The standard quantum-mechanical treatment of the oscillation phenomenon is based on the idea that the neutrino flavour state
generated in a weak interaction process is a quantum mixture of several mass eigenstates with different masses. 
The extremely fruitful quantum-mechanical approach has, however, a number of internal inconsistencies (see, e.g., Ref.~\cite{Giunti:2007ry}
and references therein) that have inspired developments towards a new approach based on $S$-matrix formalism in perturbative
quantum field theory (QFT) 
\cite{Giunti:1993se,Grimus:1996av,Mohanty:1997zi,Giunti:1997sk,Campagne:1997fu,Kiers:1997pe,Zralek:1998rp,Ioannisian:1998ch,%
      Grimus:1998uh,Grimus:1999zp,Grimus:1999ra,Cardall:1999bz,Cardall:1999ze,Beuthe:2000er,Beuthe:2001rc,Beuthe:2002ej,%
      Giunti:2002xg,Garbutt:2003ih,Asahara:2004mh,Dolgov:2005nb,Nishi:2005dc,Akhmedov:2008jn,Naumov:2009zza,Kopp:2009fa,%
      Akhmedov:2009rb,Keister:2009qn,Bernardini:2010zba,Anastopoulos:2010sf,Naumov:2010um,Akhmedov:2010ua,Akhmedov:2010ms,%
      Morris:2011sn,Akhmedov:2012uu,Akhmedov:2012mk}.
The neutrino oscillation phenomenon in this approach is nothing else than a result of interference of Feynman diagrams
which perturbatively describe the lepton number violating processes with the massive neutrino fields as internal lines (propagators)
connecting the macroscopically separated vertices of the diagram, hereafter called ``source'' and ``detector''.
Hence there is no need to use or even mention the flavor neutrino states of fields.
The external lines (legs) of the macroscopic diagrams are treated as wave packets rather than plane waves.
In order to accommodate these features, the standard $S$-matrix formalism has to be modified. In particular,
the asymptotic in- and out-states can be constructed as, e.g., superpositions of ordinary one-particle Fock states,
satisfying a set of constraints like relativistic covariance and reducibility to the Fock states in the plane-wave limit \cite{Naumov:2010um}.

In a certain approximation, the full amplitude of any process under consideration can be represented through a tensor composition
of dynamic factors describing the interactions of the external wave packets in the source and detector vertices, elements of the
lepton mixing matrix, and the following integral 
\begin{align}
\label{4dGreenFunction}
\int\frac{d^4q}{(2\pi)^4} & \,\frac{\left(\hat{q}+m\right)\mathcal{F}(q)e^{-i{qx}}}{q^2-m^2+i\epsilon} \nonumber \\
= &\ \left(i\hat{\partial}+m\right)\int\frac{d^4q}{(2\pi)^4}\,\frac{\mathcal{F}(q)e^{-i{qx}}}{q^2-m^2+i\epsilon}.
\end{align}
It is assumed that the passage to the limit $\epsilon\to0$ in Eq.~\eqref{4dGreenFunction} must be performed in a distributional sense.
Here and below $\hat{\partial}=\gamma_\mu\partial^\mu$, $\hat{q}=\gamma_\mu q^\mu$, $\gamma_\mu$ are the usual Dirac matrices
($\mu=0,1,2,3$), $m$ is the neutrino mass, $x=(T,\vec{L})$ is a 4-vector in Minkowski space representing the space-time separation
between the source and detector vertices of the macrodiagram
and, furthermore, $T$ and $|\vec{L}|$ are assumed to be large in comparison with the characteristic scales of the problem;
the scales are explicitly defined by the most probable momenta $\vec{p}_\varkappa$, masses $m_\varkappa$ and
momentum spreads $\sigma_\varkappa \ll m_\varkappa$ of the external wave packets $\varkappa$ and, implicitly
(through the dynamic factors in the full amplitude), by the local interaction Lagrangian density.
The tensor-valued function $\mathcal{F}(q)$ in Eq.~\eqref{4dGreenFunction} parametrically depends on the
configuration of the external wave packets (governed by the set $\{\vec{p}_\varkappa,m_\varkappa,\sigma_\varkappa\}$)
and is proportional to the gauge boson propagators which are independent of $\vec{q}$ in the four-fermion approximation
sufficient for the oscillation studies in the terrestrial experiments.
Hence, without loss in generality, the function $\mathcal{F}(q)$ can be thought to be a real-valued.
Within the plane-wave limit,
\[
\mathcal{F}(q)\propto\delta\left(q-q_s\right)\delta\left(q+q_d\right),
\]
where $q_s$ and $q_d$ are the 4-momentum transfers in the source and detector vertices, respectively.
So, in the general case, $\mathcal{F}(q)$ is responsible for the approximate mean energy and momentum conservation
in the vertices of the macrodiagram, with a precision again defined by the kinematic attributes of the in- and out-packets
$\{\vec{p}_\varkappa,m_\varkappa,\sigma_\varkappa\}$ and by the dynamic factors (see Ref.~\cite{Naumov:2010um} for more details).

The most interesting from the standpoint of studying the neutrino flavor oscillations is behavior of the integral
\eqref{4dGreenFunction} for macroscopically large spatial distances $L=|\vec{L}|$. 
The desired asymptotics is given by the Grimus-Stockinger (GS) theorem \cite{Grimus:1996av} which states:%
\footnote{It is assumed implicitly that the function $\varPhi$ along with its first to third derivatives are absolutely integrable
          and, moreover, that $\varPhi$ is a rotation-invariant function; the latter property will be implied from here on out.
          It is also pertinent to note that, as it can be seen from the subsequent calculations, the remainder term
          in Eq.~\eqref{GSasymptotics} is actually $\mathcal{O}(L^{-1})$, not $\mathcal{O}(L^{-1/2})$.
         }
{\theorem
\label{TheGStheorem}
Let $\varPhi(\vec{q})\in C^3(\mathbb{R}^3)$ itself and its first and second derivatives decrease at least like $1/|\vec{q}|^2$
as $|\vec{q}|\to\infty$, $\upkappa^2$ be a real number, $\vec{l}=\vec{L}/L$ and
\begin{equation}
\label{GSintegral_Origin}
J(\vec{L},\upkappa) = \int\frac{d\vec{q}}{(2\pi)^3}\,\frac{\varPhi(\vec{q})e^{-i\vec{qL}}}{\vec{q}^2-\upkappa^2-i\epsilon}.
\end{equation}
Then in the asymptotic limit $L\to\infty$ one obtains for $\upkappa>0$
\begin{equation} 
\label{GSasymptotics}
J(\vec{L},\upkappa)=\frac{e^{i\upkappa L}\varPhi(-\upkappa\vec{l})}{4{\pi}L}\left[1+\mathcal{O}\left(\frac{1}{L^{1/2}}\right)\right],
\end{equation}
whereas for $\upkappa^2<0$ the integral decreases like $L^{-2}$.
}

The physical meaning of this theorem becomes obvious after substituting Eq.~\eqref{GSasymptotics} into Eq.~\eqref{4dGreenFunction}
and a saddle-point integration in energy variable $q_0$. 
The phase factor
\[
\exp\left(i\upkappa L\right)=\exp\left(i\sqrt{q_0^2-m^2}L\right)
\]
in Eq.~\eqref{GSasymptotics} appears responsible for the oscillation behavior of the full amplitude built as a sum
of contributions with different neutrino masses, while the factor $1/L$ leads to the classical inverse-square law (ISL)
in the modulus-squared amplitude, which represents the probability and, when a proper macroscopic averaging is performed,
-- the measurable count rate of the neutrino induced events.

In Ref.~\cite{Naumov:2010um}, within the framework of the so-called contracted relativistic Gaussian packet (CRGP) model
and after imposing several restrictions on the space-time geometry of the neutrino production/detection (gedanken)
experiment, the following result for the neutrino event rate has been derived:
\begin{equation}
\label{EventRate}
\frac{dN}{d\tau}
 = {V_{\mathcal{D}}V_{\mathcal{S}}}\int_{V_{\mathcal{S}}} d\vec{x}\int_{V_{\mathcal{D}}} d\vec{y}\int d\mathfrak{F}_\nu
   \int d\sigma_{{\nu}\mathcal{D}}\mathcal{P}_{\alpha\beta}\left(E_\nu,\left|\vec{y}-\vec{x}\right|\right).
\end{equation}
Here $\tau$ is the detector exposure time, $E_\nu$ is the neutrino energy, $V_{\mathcal{S}}$ and $V_{\mathcal{D}}$ are
the spatial volumes of, respectively, the source $\mathcal{S}$ and detector $\mathcal{D}$ ``devices''.%
\footnote{More accurately, $\mathcal{S}$ and $\mathcal{D}$ are the supports of the products of the one-particle distribution
          functions $f_a(\vec{p}_a,\vec{x}_a)$ of the in-particles $a$, expressed in terms of the most probable momenta
          $\vec{p}_a$ and spatial coordinates, $\vec{x}_a$, of the centers of the external wave packets.
          It is supposed for simplicity that the functions $f_a$ are time-independent during the detection period,
          the devices $\mathcal{S}$ and $\mathcal{D}$ are finite and mutually disjoint within the space domain,
          and their effective spatial dimensions are small compared to the mean distance between them but very large
          compared to the effective dimensions ($\sim\sigma_\varkappa^{-1}$) of all in and out wave packets moving inside
          $\mathcal{S}$ and $\mathcal{D}$ (see Ref.~\cite{Naumov:2010um} for more details).
         }
The differential form $d\sigma_{{\nu}\mathcal{D}}$ is defined as that the expression
\begin{equation*}
\label{CrossSection}
\frac{1}{V_{\mathcal{D}}}\int d\vec{y}d\sigma_{{\nu}\mathcal{D}}
\end{equation*}
represents the differential cross section of the neutrino scattering off the detector $\mathcal{D}$ as a whole.
In the particular and the most practically important case of neutrino scattering from single particles,
provided that the momentum distribution of the scatterers is sufficiently narrow (Maxwellian distribution is a good
example), the differential form $d\sigma_{{\nu}\mathcal{D}}$ becomes exactly the elementary differential cross section
of this reaction multiplied by the total number of the scatterers in the detector volume.
The factor $\mathcal{P}_{\alpha\beta}(E_\nu,L)$ in Eq.~\eqref{EventRate}
is the QFT generalization of the quantum-mechanical flavor transition probability.%
\footnote{As is shown in Ref.~\cite{Naumov:2010um}, the factor $\mathcal{P}_{\alpha\beta}(E_\nu,L)$ is not really the probability
          since it does not in general satisfy the unitarity relation. However, this fact is not essential in the context
          of the present study and this quantity can be referred to as the probability.}
Finally, the differential form $d\mathfrak{F}_\nu$ is defined in such a way that the quantity
\begin{equation*}
\label{NeutrinoFlux}
\frac{d\vec{x}}{V_{\mathcal{S}}}\int \frac{d\mathfrak{F}_\nu}{dE_{\nu}}
\end{equation*}
be the flux density of neutrinos in the detector, or, more precisely, the number of neutrinos appearing per unit time
and unit neutrino energy in an elementary volume $d\vec{x}$ around the point $\vec{x}\in\mathcal{S}$, 
travelling within the solid angle $d\vec{l}$ about the flow direction $\vec{l}=(\vec{y}-\vec{x})/|\vec{y}-\vec{x}|$
and crossing a unit area, placed around the point $\vec{y}\in\mathcal{D}$ and normal to the vector $\vec{l}$.
As a consequence of the GS theorem,
\begin{equation}
\label{TheISL}
d\mathfrak{F}_\nu \propto |\vec{y}-\vec{x}|^{-2},
\end{equation}
that results just to the classical ISL for the neutrino flux.
 
The asymptotic nature of Eq.~\eqref{GSasymptotics} itself suggests that the relation \eqref{TheISL} may break down
at short distances, but the theorem~\ref{TheGStheorem} does not provide the physical scale for this breakdown,
$\mathfrak{L}_0=\mathfrak{L}_0(\vec{l},\upkappa)$.
It is stated in Ref.~\cite{Grimus:1996av} that ``large $L$'' means
\begin{equation}
\label{TheGScondition}
\overline{E}_{\nu}L \approx 5\times10^{10}\left(\frac{\overline{E}_{\nu}}{1\,\text{MeV}}\right)
                                          \left(\frac{L}{1\,\text{cm}}\right) \gg 1,
\end{equation}
where $\overline{E}_{\nu}$ is an average (anti)neutrino energy. This condition is apparently well satisfied
for every thinkable neutrino experiment. But, inasmuch as $|J(\vec{L},\upkappa)|$ ceases to depend on $L$
in the plane-wave limit (that is, at $\sigma_\varkappa=0$, $\forall\varkappa$), it may be inferred that
$\mathfrak{L}_0$ is affected not only by the neutrino energy, but also by the momentum spreads of the external
wave packets; hence the very mild condition \eqref{TheGScondition} is generally not sufficient. 
A certain degree of information on the scale parameter $\mathfrak{L}_0$ can be extracted from the corrections
to the GS asymptotics \eqref{GSasymptotics} derived in terms of powers of $1/L$.
Indeed, according to Refs.~\cite{Akhmedov:2010ms,Korenblit:2013tya}, the leading-order relative correction to the
GS asymptotics is proportional to $\upkappa/(\sigma_{\text{eff}}^2L)$, where $\sigma_{\text{eff}}$ is a representative
scale of the external momentum spreads.
In Refs.~\cite{Akhmedov:2010ms,Korenblit:2013tya}, it is however assumed that the power corrections are negligible under
the conditions of all neutrino-oscillation experiments. But in fact the scale
\begin{equation}
\label{L0}
\mathfrak{L}_0 \sim      \upkappa\sigma_{\text{eff}}^{-2}
    \approx 20 \left(\frac{\upkappa}{1\;\text{MeV}}\right)
               \left(\frac{\sigma_{\text{eff}}}{1\;\text{eV}}\right)^{-2}\;\text{cm}
\end{equation}
can be macroscopically large at sufficiently high energies and/or small momentum spreads, thereby indicating on either inapplicability
of the asymptotic expansion in powers of $1/L$ or on the \emph{real} possibility of the ISL violation at $L\lesssim \mathfrak{L}_0$,
which could, in principle, be measurable.
In particular, it is not improbable that the ISL violation can be relevant to the observed deficit of $\nu_e$ and
$\overline{\nu}_e$ interaction rates (compared to the expected values) in the calibration experiments with artificial
neutrino sources in the Ga-Ge radiochemical detectors (``Gallium neutrino anomaly'' \cite{Giunti:2010zu}) and/or
in the short-baseline (SBL) reactor experiments (``Reactor antineutrino anomaly'' \cite{Mention:2011rk,Mueller:2011nm,Huber:2011wv}).
Therefore, it is interesting to study this issue further.

In the next section, we prove a theorem which yields the leading power correction to the GS asymptotics,
by using a straightforward extension of the original method by Grimus and Stockinger.
Since, as it will be seen, this correction is (or, generally, can be) purely imaginary, it itself cannot provide
the necessary conditions for the ISL violation.
The GS method can certainly be used, with successive narrowing of the class of the functions $\varPhi(\vec{q})$,
for deriving the higher-order corrections, but it becomes too tedious for this purpose.
That is why in section~\ref{sect:LessRigorousApproach} we apply a much simpler, while less rigorous method,
a generalization of that used in Ref.~\cite{Korenblit:2013tya}, which allows us to calculate the asymptotic corrections
to any order in $1/L$ for both $J(\vec{L},\upkappa)$ and $|J(\vec{L},\upkappa)|^2$.
Finally, in section~\ref{sect:TheISLviolation}, we discuss the potential relevance of the ISL violation
to the observed gallium and reactor anomalies.


\section{Extended Grimus-Stockinger theorem}
\label{sect:ExtendedGStheorem}

{\theorem
\label{ExtendedGStheorem}
Let $\varPhi(\vec{q})\in C^4(\mathbb{R}^3)$ itself and its first to third derivatives decrease at least like
$1/|\vec{q}|^3$ as $|\vec{q}|\to\infty$. Let, besides, $\bnabla_{\vec{q}}^m\varPhi(\vec{q})$, $m=0,\ldots,4$
be absolutely integrable and $\upkappa^2$ be a real number.
Then in the asymptotic limit $L\to\infty$ one obtains for $\upkappa>0$ 
\begin{align}
\label{ExtendedGStheorem-Part-I}
 J(\vec{L},\upkappa)
= &\  \frac{e^{i\upkappa L}}{4{\pi}L}
      \left\{1-\frac{i}{L}\left[\left(\vec{l}\bnabla_{\vec{q}}\right)
     +\frac{\upkappa}{2}\left(\vec{l}\times\!\bnabla_{\vec{q}}\right)^2\right] \right. \nonumber \\
  &\ \left.\left.+\mathcal{O}\left(\frac{1}{L^{2}}\right)\right\}\varPhi(\vec{q})\right|_{\vec{q}=-\upkappa\vec{l}},
\end{align}
whereas for $\upkappa^2<0$ the integral decreases like $L^{-4}$.
}

\proof

The second part of the theorem is just a corollary of the following Lemma:
{\lemma
\label{3dLocalizationPrinciple}
Suppose $\Psi(\vec{q})\in C^4(\mathbb{R}^3)$, $\bnabla_{\vec{q}}^m\Psi(\vec{q})$, $m=0,\ldots,4$
are absolutely integrable
and $\bnabla_{\vec{q}}^m\Psi(\vec{q})$, $m=0,1,3$
vanish on the boundary $\partial{Q}$ of a simply connected domain $Q$
(if $Q=\mathbb{R}^3$ then the functions $\bnabla_{\vec{q}}^m\Psi(\vec{q})$
must decay faster than $1/|\vec{q}|^3$ as $|\vec{q}|\to\infty$).
Then the function
\begin{equation}
\label{Int_Psi}
\mathcal{J}(\vec{L})=\left|\int_Q d\vec{q}\Psi(\vec{q})e^{-i\vec{L}\vec{q}}\right|
\end{equation}
decreases faster than $L^{-4}$ as $L\to\infty$.
}

According to Green's theorem, 
for the functions $\Psi(\vec{q})$ and $\Phi(\vec{q})$ having continuous partial derivatives of the second order
the following identity holds:
\begin{align*}
\label{3dLocalizationPrinciple-1}
 \int_Q & d\vec{q}\left[\Psi(\vec{q})\bnabla_{\vec{q}}^2\Phi(\vec{q})-\Phi(\vec{q})\bnabla_{\vec{q}}^2\Psi(\vec{q})\right]  \\
        & = \int_{\partial{Q}} d\vec{S}\left[\Psi(\vec{q})\bnabla_{\vec{q}}\Phi(\vec{q})-\Phi(\vec{q})\bnabla_{\vec{q}}\Psi(\vec{q})\right].
\end{align*}
Substituting $\Phi(\vec{q})=e^{-i \vec{L} \vec{q}}$ and, considering that $\Psi(\vec{q})=0$ and $\bnabla_{\vec{q}}\Psi(\vec{q})=0$
as $\vec{q}\in\partial Q$, yields
\begin{equation}
\label{3dLocalizationPrinciple-2}
-{L^2}\int_Qd\vec{q}\Psi(\vec{q})e^{-i\vec{L}\vec{q}} = \int_Qd\vec{q}e^{-i\vec{L}\vec{q}}\bnabla_{\vec{q}}^2\Psi(\vec{q}).
\end{equation}
Similarly, we can write
\begin{equation*}
\int_Q d\vec{q}\left[\bnabla_{\vec{q}}^2\Psi(\vec{q})\right]\left[\bnabla_{\vec{q}}^2\Phi(\vec{q})\right] 
= \int_Q d \vec{q}\Phi(\vec{q})\bnabla_{\vec{q}}^2\left[\bnabla_{\vec{q}}^2\Psi(\vec{q})\right],
\end{equation*}
or, in our particular case,
\begin{equation}
\label{3dLocalizationPrinciple-3}
-L^2\int_Qd\vec{q}\bnabla_{\vec{q}}^2\Psi(\vec{q})e^{-i\vec{Lq}}
=   \int_Qd\vec{q}e^{-i \vec{L}\vec{q}}\bnabla_{\vec{q}}^4\Psi(\vec{q}).
\end{equation}
Combining Eqs.~\eqref{3dLocalizationPrinciple-2} and \eqref{3dLocalizationPrinciple-3} together yields the statement of the Lemma.


{\remark 
By repeating the steps of the proof one can show that the integral \eqref{Int_Psi} decays faster than any power of $1/L$
if $\Psi(\vec{q})\in C^\infty_0(\mathbb{R}^3)$ is a function that decreases at infinity together with all its derivatives
faster than $1/|\vec{q}|^3$.
}

From here on we will assume that $\upkappa>0$.
Following the approach of Grimus and Stockinger \cite{Grimus:1996av} we divide the proof of the first part of
the theorem into several steps. Let us start with the following Lemma:


{\lemma
Let $f(\theta) \in C^4([0,\pi])$ 
and
\begin{equation} 
\label{GSintegral_in_theta}
I(r) = \int_0^{\pi}d\theta\sin\theta\,f(\theta)e^{-ir\cos\theta}.
\end{equation}
Then the following asymptotic expansion holds true as $r\to\infty$, $r\in\mathbb{R}$:
\begin{align} 
\label{GSlemma_1a}
I(r) = &\  \frac{1}{ir}\left[e^{ir}f(\pi)-e^{-ir}f(0)\right] \nonumber \\
       &\ +\sqrt{\frac{\pi}{2r^3}}\left[f'(\pi)e^{ir_+}-f'(0)e^{-ir_+}\right] \nonumber \\
       &\ +\frac{1}{r^2}\left[e^{ir}f_{\pi}(0)-e^{-ir}f_0(0)\right] \nonumber \\
       &\ +\sqrt{\frac{\pi}{2r^5}}\left\{e^{ ir_-}\left[\frac{1}{8}f'(\pi)+f_{\pi}'(0)\right]\right. \nonumber \\
       &\ \left.                        -e^{-ir_-}\left[\frac{1}{8}f'(0)  +f_0'    (0)\right]\right\}
        +\mathcal{O}\left(\frac{1}{r^3}\right),
\end{align}
where $r_{\pm}=r\pm\pi/4$, $f_{\pi}(\theta)=\left[f'(\pi-\theta)-f'(\pi)\right]/{\sin\theta}$ and
$f_0(\theta)=\left[f'(\theta)-f'(0)\right]/{\sin\theta}$.
}

Integrating Eq.~\eqref{GSintegral_in_theta} by parts, we obtain
\begin{align} 
\label{IntegratingByParts}
I(r)   = &\ \frac{i}{r}\left[e^{-ir}f(0)-e^{ir}f(\pi)+I_1(r)\right], \nonumber \\
I_1(r) = &\ \int_0^{\pi}d\theta e^{-ir\cos\theta}f'(\theta).
\end{align}
Let us split $I_1(r)$ into two parts:
\begin{equation}
\label{I_1inTwoParts}
I_1(r)=I_{11}(r)+I_{12}(r),
\end{equation}
where
\begin{align*}
I_{11}(r) = &\ \int_0^{\pi/2}d\theta e^{-ir\cos\theta}f'(\theta), \\
I_{12}(r) = &\ \int_0^{\pi/2}d\theta e^{+ir\cos\theta}f'(\pi-\theta).
\end{align*}
By using definition of the function $f_0(\theta)$, we get
\begin{align}
\label{I_11}
I_{11}(r) = &\ f'(0)\int_0^{\pi/2}d\theta e^{-ir\cos\theta} \nonumber \\
            &\ +\int_0^{\pi/2}d\theta\sin\theta\,f_0(\theta)e^{-ir\cos\theta}.
\end{align}
According to one of the definitions of the Bessel and Struve functions
\begin{equation}
\label{J_0}
J_0(z)          = \frac{2}{\pi}\int_0^{\pi/2}d\theta\cos(z\cos\theta)
\end{equation}
and
\begin{equation}
\label{H_0}
\mathbf{H}_0(z) = \frac{2}{\pi}\int_0^{\pi/2}d\theta\sin(z\cos\theta),
\end{equation}
we can write
\begin{align}
\label{GrimIntegral}
\int_0^{\pi/2}d\theta e^{-ir\cos\theta} 
= &\ \frac{\pi}{2}\left[J_0(r)-i\mathbf{H}_0(r)\right] \nonumber \\
= &\ \sqrt{\frac{\pi}{2r}}e^{-ir_-}-\frac{i}{r}        \nonumber \\               
  &\ +\frac{i}{8}\sqrt\frac{\pi}{2r^3}e^{-ir_-}+\mathcal{O}\left(\frac{1}{r^2}\right),               
\end{align}
where the well-known formulas for the asymptotics of the functions \eqref{J_0} and \eqref{H_0} were used.
Consider now the second integral in Eq.~\eqref{I_11}.
Since the function $f_0(\theta)$ is thrice differentiable, we can employ integration
by parts, which yields
\begin{align}
\label{SecondIntegral}
\int_0^{\pi/2}d\theta\sin\theta\,f_0(\theta)e^{-ir\cos\theta}
= &\ \frac{i}{r}\left[e^{-ir}f_0(0)-f_0\left(\frac{\pi}{2}\right)\right. \nonumber \\
  &\ \left.+\int_0^{\pi/2}d\theta e^{-ir\cos\theta}f_0'(\theta)\right].
\end{align}
By introducing the function $\xi(\theta)=\left[f_0'(\theta)-f_0'(0)\right]/\sin\theta$,
the right hand part of the last equality can be rewritten as follows:
\begin{align}
\label{SecondIntegrationIn_theta}
\frac{i}{r}&\left[e^{-ir}f_0(0)-f_0\left(\frac{\pi}{2}\right)\right] \nonumber \\
&\ +\frac{i}{r}\int_0^{\pi/2}d\theta\left[f_0'(0)e^{-ir\cos\theta}+\sin\theta e^{-ir\cos\theta}\xi(\theta)\right].
\end{align}
As the function $\xi(\theta)$ is twice differentiable, we can apply integration by parts to the
corresponding part of the integral in Eq.~\eqref{SecondIntegrationIn_theta}, after which it becomes obvious that this
part decreases like $\mathcal{O}(r^{-2})$.
So, having regard to Eq.~\eqref{GrimIntegral}, the integral \eqref{SecondIntegral} is equal to
\begin{equation}
\label{NeededForI2_r}
\frac{i}{r}\left[e^{-ir}f_0(0)-f_0\left(\frac{\pi}{2}\right)\right]
  +if_0'(0)\sqrt{\frac{\pi}{2r^3}}e^{-ir_-} 
 +\mathcal{O}\left(\frac{1}{r^2}\right).           
\end{equation}
Substituting Eqs.~\eqref{NeededForI2_r} and \eqref{GrimIntegral} into Eq.~\eqref{I_11} then yields
\begin{align} 
\label{I_11_a}
I_{11}(r)
= &\ f'(0)\sqrt{\frac{\pi}{2r}}e^{-ir_-}+\frac{1}{ir}\left[f'(0)+f_0\left(\frac{\pi}{2}\right)-e^{-ir}f_0(0)\right] \nonumber \\
  &\ +i\sqrt{\frac{\pi}{2r^3}}e^{-ir_-}\left[\frac{1}{8}f'(0)+f_0'(0)\right]+\mathcal{O}\left(\frac{1}{r^2}\right).           
\end{align}
Taking complex-conjugate of Eq.~\eqref{GrimIntegral} yields
\begin{equation}
\label{GrimIntegral_compl_sop}
\int_0^{\pi/2}d\theta e^{ir\cos\theta}= \sqrt{\frac{\pi}{2r}}e^{ir_-}+\frac{i}{r} 
          -\frac{i}{8}\sqrt{\frac{\pi}{2r^3}}e^{ir_-}+\mathcal{O}\left(\frac{1}{r^2}\right).         
\end{equation}
Now we can rewrite the integral $I_{12}(r)$ in terms of the thrice differentiable function $f_{\pi}(\theta)$:
\begin{equation} 
\label{I_12}
I_{12}(r)=f'(\pi)\int_0^{\pi/2}d\theta e^{ir\cos\theta}+\int_0^{\pi/2}d\theta e^{ir\cos\theta}\sin\theta f_{\pi}(\theta).
\end{equation}
Integrating the second integral in Eq.~\eqref{I_12} by parts, we get
\begin{align}
\label{second_part_with_dividing_1}
\int_0^{\pi/2}d\theta\sin\theta f_{\pi}(\theta)e^{ir\cos\theta}
= &\ \frac{i}{r}\left[f_{\pi}\left(\frac{\pi}{2}\right)-e^{ir}f_{\pi}(0)\right] \nonumber \\
  &\ -\frac{i}{r}\int_0^{\pi/2}d\theta e^{ir\cos\theta}f_{\pi}'(\theta).
\end{align}
The expression on the right-hand part of Eq.~\eqref{second_part_with_dividing_1} can be written in the form
\begin{align}
\label{second_part_with_dividing_2}
\frac{i}{r}&\left[f_{\pi}\left(\frac{\pi}{2}\right)-e^{ir}f_{\pi}(0)\right] \nonumber \\
&\ -\frac{i}{r}\int_0^{\pi/2}d\theta\left[f_{\pi}'(0)e^{ir\cos\theta}+\sin\theta\lambda(\theta)e^{ir\cos\theta}\right],
\end{align}
where $\lambda(\theta)=\left[f_{\pi}'(\theta)-f_{\pi}'(0)\right]/\sin\theta$.
Since $\lambda(\theta)$ is twice differentiable function, one can apply a partial integration to the
second integral in Eq.~\eqref{second_part_with_dividing_2}.
This shows that integral \eqref{second_part_with_dividing_1} decreases like $\mathcal{O}(r^{-2})$.
Now, by substituting Eq.~\eqref{GrimIntegral_compl_sop} into the previous expression yields
\begin{align*}
\int_0^{\pi/2} & d\theta\sin\theta\,f_{\pi}(\theta)e^{ir\cos\theta} \\
= &\  \frac{i}{r}\left[f_{\pi}\left(\frac{\pi}{2}\right)-e^{ir}f_{\pi}(0)\right] 
     -if_{\pi}'(0)\sqrt{\frac{\pi}{2r^3}}e^{ir_-}+\mathcal{O}\left(\frac{1}{r^2}\right)    
\end{align*}
and, taking into account Eqs.~\eqref{GrimIntegral_compl_sop} and \eqref{I_12}, we obtain
\begin{align}
\label{I_12_b}
I_{12}(r) = &\ \sqrt{\frac{\pi}{2r}}f'(\pi)e^{ir_-}+\frac{i}{r}\left[f'(\pi)+f_{\pi}\left(\frac{\pi}{2}\right)-e^{ir}f_{\pi}(0)\right] \nonumber \\
            &\  -i\sqrt{\frac{\pi}{2r}}e^{ir_-}\left[\frac{1}{8}f'(\pi)+f_{\pi}'(0)\right]+\mathcal{O}\left(\frac{1}{r^2}\right).                        
\end{align}
By substituting Eqs.~\eqref{I_11_a} and \eqref{I_12_b} into Eq.~\eqref{I_1inTwoParts} and using the
explicit form of the functions $f_0(\theta)$ and $f_{\pi}(\theta)$, we arrive at the following result:
\begin{align}
\label{CompletingProofOfTheLemma} 
I_1(r) 
= &\ \sqrt{\frac{\pi}{2r}}\left[f'(0)e^{-ir_-}+f'(\pi)e^{ir_-}\right] \nonumber \\
  &\ +\frac{i}{r}\left[e^{-ir}f_0(0)-e^{ir}f_{\pi}(0)\right]          \nonumber \\
  &\ +i\sqrt{\frac{\pi}{2r^3}}\left\{e^{-ir_-}\left[\frac{1}{8}f'(0)+f_0'(0)\right]\right.  \nonumber \\
  &\ \left.-e^{ir_-}\left[\frac{1}{8}f'(\pi)+f_{\pi}'(0)\right]\right\}+\mathcal{O}\left(\frac{1}{r^2}\right).
\end{align}
Finally, by substituting Eq.~\eqref{CompletingProofOfTheLemma} into Eq.~\eqref{IntegratingByParts} we complete the proof of the Lemma.

From this Lemma it immediately follows the asymptotic expansions of the integral
\begin{align}
\label{corollary1}
I_w(r) \equiv &\ \int_0^{\pi}d\theta\sin\theta\sin(wr\cos\theta)\,f(\theta)e^{-ir\cos\theta} \nonumber \\
          =   &\ \frac{1}{2i}\left[I(r-wr)-I(r+wr)\right]
\end{align}
valid for $|w|<1$.

Following Ref.~\cite{Grimus:1996av} we write $J=J_1+J_2$, where
\begin{align*}
J_1 = &\ \frac{i\upkappa}{(4\pi)^2}\int_{S^2} d\upkappa\varPhi(\upkappa\vec{n})e^{-i\upkappa \vec{n}\vec{L}}, \\
J_2 = &\ \int\frac{d\vec{q}}{(2\pi)^3}\varPhi(\vec{q})e^{-i\vec{qL}}\frac{q^2-\upkappa^2}{(q^2-\upkappa^2)^2+\epsilon^2},
\end{align*}
$S^2$ is the 2-dimensional unit sphere, $q\equiv|\vec{q}|$ and $\vec{n}\equiv\vec{q}/q$.
Our objective now is to derive the asymptotic behavior of the integrals $J_1$ and $J_2$.

\subsection{Integral \protect$J_1(\vec{L},\upkappa)$}
\label{sect:J_1}

Without loss of generality, we can use the coordinate frame in which $\vec{l} = \left(0,0,1\right)$
and $\vec{n} = \left(\sin\theta\cos\varphi,\sin\theta\sin\varphi,\cos\theta\right)$.
Then, by applying Lemma \ref{3dLocalizationPrinciple} with $r=\upkappa L$ we arrive at the asymptotic expansion
\begin{align*}
J_1(\vec{L},\upkappa)
= &\  \frac{g_1(L,\upkappa)}{2{\pi}\upkappa L} 
     -\frac{g_{3/2}(L,\upkappa)}{\pi(2\pi\upkappa L)^{3/2}}
     +\frac{ig_2(L,\upkappa)}{(2\pi\upkappa L)^2} \\
  &\ -\frac{i{\pi}g_{5/2}(L,\upkappa)}{(2\pi\upkappa L)^{5/2}}
     +\mathcal{O}\left(\frac{1}{L^3}\right),
\end{align*}
in which
\def\VPHANTOM{\vphantom{\frac{1}{8}}}
\begin{align*}
g_1(L,\upkappa)     = &\ \frac{\upkappa}{4}\left[e^{i\upkappa L}\varPhi(\upkappa,\varphi,\pi)-e^{-i\upkappa L}\varPhi(\upkappa,\varphi,0)\right],      \\
g_{3/2}(L,\upkappa) = &\ \frac{\upkappa}{4}\int_0^{2\pi}d\varphi
                         \left[e^{-i\rho}\left.\partial_{\theta}\varPhi(\upkappa,\varphi,\theta)\right|_{\theta=0}     \right.                         \\  
                      &\ \left.+e^{ i\rho}\!\left.\partial_{\theta}\varPhi(\upkappa,\varphi,\theta)\right|_{\theta=\pi}\right],                        \\
g_2(L,\upkappa)     = &\ \frac{\upkappa}{4}\int_0^{2\pi}d\varphi\left[e^{ i\upkappa L}\varPhi_{\pi}(\upkappa,\varphi,0)\right.                         \\
                      &\ \left.                                      -e^{-i\upkappa L}\varPhi_0    (\upkappa,\varphi,0)\right],                        \\
g_{5/2}(L,\upkappa) = &\ \frac{\upkappa}{4}\int_0^{2\pi}d\varphi
                         \left[e^{-i\rho}\left(\frac{1}{8}\left.\partial_{\theta}\varPhi(\upkappa,\varphi,\theta)\right|_{\theta=0}\right.\right.      \\
                      &\ \left.\left.  \left.+\partial_{\theta}\varPhi_0(\upkappa,\varphi,\theta)\right|_{\theta=0}\VPHANTOM \right)\right.            \\
                      &\ \left. -e^{ i\rho}\left(\frac{1}{8}\left.\partial_{\theta}\varPhi(\upkappa,\varphi,\theta)\right|_{\theta=\pi} \right.\right. \\
                      &\ \left.\left. +\left.  \partial_{\theta}\varPhi_{\pi}(\upkappa,\varphi,\theta)\right|_{\theta=0}\VPHANTOM\right)\right].
\end{align*}
Here we introduced the notation
\begin{align*}
\varPhi_0(\upkappa,\varphi,\theta)
                    = &\  \frac{\partial_\theta\varPhi(\upkappa,\varphi,\theta)
                         -\left.\partial_\theta\varPhi(\upkappa,\varphi,\theta)\right|_{\theta=0}}{\sin\theta}, \\
\varPhi_{\pi}(\upkappa,\varphi,\theta) 
                    = &\ \frac{\left.\partial_{\vartheta}\varPhi(\upkappa,\varphi,\vartheta)\right|_{\vartheta=\pi-\theta}
                              -\left.\partial_{\theta   }\varPhi(\upkappa,\varphi,\vartheta)\right|_{\vartheta=\pi}}{\sin\theta};
\end{align*}
$\rho=\upkappa L-\pi/4$ and $\varPhi(\upkappa,\varphi,\theta)\equiv\varPhi(\upkappa\vec{n})$.
Let us prove that
\begin{equation}
\label{g_n/2}
g_{3/2}(L,\upkappa)=g_{5/2}(L,\upkappa)=0.
\end{equation}
For this purpose we consider the integral
\begin{align}
\label{NullIntegral}
\int_0^{2\pi} & d\varphi\left.\partial_{\theta}\varPhi(\upkappa,\varphi,\theta)\right|_{\theta=0}                                               \nonumber \\
& = \int_0^{2\pi}d\varphi\left.\frac{\partial\varPhi(\upkappa\vec{n})}{\partial\vec{n}}\frac{\partial\vec{n}}{\partial\theta}\right|_{\theta=0} \nonumber \\
& = \int_0^{2\pi}d\varphi\left[\frac{\partial\varPhi(\upkappa\vec{n})}{\partial n_x}\cos\varphi
   +\frac{\partial\varPhi(\upkappa\vec{n})}{\partial n_y}\sin\varphi\right]_{\theta=0}.
\end{align}
Since $n_x=n_y=0$ as $\theta=0$, the functions $\left[\partial\varPhi(\upkappa\vec{n})/\partial n_i\right]_{\theta=0}$ ($i=x,z$) depend only on $\upkappa$
and $n_z$ and thus do not depend on $\varphi$ and can be factored out the integral.
So the integral \eqref{NullIntegral} is equal to zero. Similar consideration shows that 
\begin{equation*}
\int_0^{2\pi}d\varphi\left.\partial_{\theta}\varPhi(\upkappa,\varphi,\theta)\right|_{\theta=\pi} = 0.
\end{equation*}
Next, it can be seen that
\begin{equation*}
\int_0^{2\pi}d\varphi \left.\partial_{\theta}\varPhi_0(\upkappa,\varphi,\theta)\right|_{\theta=0} = 0.
\end{equation*}
Indeed, by l'Hospital's rule
\begin{equation*}
 \left.\partial_{\theta}\varPhi_0(\upkappa,\varphi,\theta)\right|_{\theta=0}
= \frac{1}{2}\left[\partial^3_{\theta}\varPhi(\upkappa,\varphi,\theta)
+\partial_{\theta}\varPhi(\upkappa,\varphi,\theta)\right]_{\theta=0}.
\end{equation*}
We already proved that the integral in $\varphi$ of the function $\left.\varPhi(\upkappa,\varphi,\theta)\right|_{\theta=0}$ is equal to zero.
In similar way, it can be proved that the integrals of the functions
$\left.\partial^3_{\theta}\varPhi(\upkappa,\varphi,\theta)\right|_{\theta=0}$
and 
$\left.\partial_{\theta}\varPhi_{\pi}(\upkappa,\varphi,\theta)\right|_{\theta=0}$
vanish too. Hence the equations \eqref{g_n/2} are proved and, as a result, we obtain 
\begin{align*}
\label{J_1}
J_1(\vec{L},\upkappa)
= &\  \frac{1}{8{\pi}L} \left[e^{i\upkappa L}\varPhi(\upkappa,\varphi,\pi)-e^{-i\upkappa L}\varPhi(\upkappa,\varphi,0)\right]  \\
  &\ +\frac{i\upkappa}{(4\pi{\upkappa}L)^2}\int_0^{2\pi}d\varphi\left[e^{ i\upkappa L}\varPhi_{\pi}(\upkappa,\varphi,0)\right. \\
  &\  \left.                                                         -e^{-i\upkappa L}\varPhi_0    (\upkappa,\varphi,0)\right] 
     +\mathcal{O}\left(\frac{1}{L^3}\right).
\end{align*}

\subsection{Integral \protect$J_2(\vec{L},\upkappa)$}
\label{sect:J_2}

According to Ref.~\cite{Grimus:1996av}, the integral $J_2$ can be split into three parts:
$J_2 = J_{21}+J_{22}+J_{23}$,
where
\begin{align*}
J_{21} = &\ \int\frac{d\vec{q}}{(2\pi)^3}e^{-i\vec{qL}}
                            \left[\varPhi(\vec{q})-\varPhi(\upkappa \vec{n})h(q-\upkappa)\right]
                            \chi_1(q,\upkappa), \\
J_{22} = &\ \int\frac{d\vec{q}}{(2\pi)^3}e^{-i\vec{qL}}\varPhi(\upkappa \vec{n})\left[\chi_1(q,\upkappa)-\frac{\chi_2(q,\upkappa)}{q^2}\right], \\
J_{23} = &\ \int_{S^2}\frac{d\upkappa}{(2\pi)^3}\int_0^{\infty}dq e^{-i\vec{qL}}\varPhi(\upkappa\vec{n})h(q-\upkappa)\chi_2(q,\upkappa);
\end{align*}
\begin{equation*}
\chi_1(q,\upkappa) = \frac{q^2-\upkappa^2}{(q^2-\upkappa^2)^2+\epsilon^2},
\quad
\chi_2(q,\upkappa) = \frac{2\upkappa^3(q-\upkappa)}{4\upkappa^2(q-\upkappa)^2+\epsilon^2},
\end{equation*}
and $q=|\vec{q}|$. The function $h(v)\in C_0^\infty(\mathbb{R})$ appearing here is a real-valued even function%
\footnote{A simple example of such a function at  $v\in\text{supp}(h)$ 
          is
          \[
          h(v) = \exp\left\{4-\left(1-a\eta+a|v|\right)^{-2}\left[1-\left(1-a\eta+a|v|\right)^2\right]^{-1}\right\},
          \]
          where $a=(1-1/\sqrt{2})/(\eta-\delta)>0$.
         }
such that $0 \leq h(v) \leq 1$, $\forall v\in\mathbb{R}$, $h(v)=1$ for $|v|\leq\delta$ and $h(v)=0$ for $|v|\geq\eta$,
where $\delta$ and $\eta$ are real parameters (having dimension of energy) which satisfy the inequalities
$0<\delta<\eta<\upkappa$.
It has been shown in Ref.~\cite{Grimus:1996av} that, under the conditions of theorem~\ref{TheGStheorem}, the
functions $J_{21}$ and $J_{22}$ decrease like $L^{-2}$. With Lemma~\ref{3dLocalizationPrinciple} it is easy to verify
that, under the conditions of theorem~\ref{ExtendedGStheorem}, these functions decrease like $L^{-4}$, and so only
the term $J_{23}$ contributes into the integral $J_2$.
Consequently, according to Eqs.~\eqref{corollary1} with $w=v/\upkappa$, $|w|<1$ and \eqref{GSintegral_in_theta},
the preasymptotic behavior of $J_{23}$ is given by
\begin{align*}
J_{23}(\vec{L},\upkappa) 
= &\  \frac{ij_1}{(2\pi)^2{\upkappa}L} 
     +\frac{1}{(2\pi)^{5/2}}\left[\frac{ij_{3/2}}{(\upkappa L)^{3/2}}-\frac{j_{5/2}}{(\upkappa L)^{5/2}}\right] \\
  &\ +\frac{j_2}{(2\pi)^3({\upkappa}L)^2}+\mathcal{O}\left(\frac{1}{L^3}\right),
\end{align*}
where
\begin{align*}
j_1        = &\  \frac{1}{2}\int_{-\eta}^\eta dv\left[ A_1(v)  \varPhi(\upkappa,\varphi,\pi) 
                                                     -A_1^*(v)\varPhi(\upkappa,\varphi,0  )\right],                                      \\
j_{3/2}    = &\  \frac{1}{2}\int_{-\eta}^{\eta}dv\int_0^{2\pi}d\varphi
                 \left[ A_{3/2}  (v)\left.\partial_{\theta}\varPhi(\upkappa,\varphi,\theta)\right|_{\theta=0  }\right.                   \\
             &\  \left.+A_{3/2}^*(v)\left.\partial_{\theta}\varPhi(\upkappa,\varphi,\theta)\right|_{\theta=\pi}\right],                  \\
j_{5/2}    = &\  \frac{1}{2}\int_{-\eta}^{\eta}dv\int_0^{2\pi}d\varphi 
                 \Big\{A_{5/2}(v)  \left[\left.\partial_{\theta}\varPhi  (\upkappa,\varphi,\theta)\right|_{\theta=0}\right.              \\
             &\  \left.                 +\left.\partial_{\theta}\varPhi_0(\upkappa,\varphi,\theta)\right|_{\theta=0}\right]           
                      -A_{5/2}^*(v)\left[\left.\partial_{\theta}\varPhi      (\upkappa,\varphi,\theta)\right|_{\theta=\pi}\right.        \\
             &\  \left.                 +\left.\partial_{\theta}\varPhi_{\pi}(\upkappa,\varphi,\theta)\right|_{\theta=0  }\right]\Big\}, \\
j_2        = &\  \frac{1}{2}\int_{-\eta}^{\eta}dv\int_0^{2\pi}d\varphi\left[ A_2(v)  \varPhi_0    (\upkappa,\varphi,0) \right.           \\
             &\  \left.                                                     -A_2^*(v)\varPhi_{\pi}(\upkappa,\varphi,0)\right];
\end{align*}
\begin{align*}
A_1(v)     = &\  e^{ i\upkappa L}\left[\cos(vL)-i\frac{\upkappa}{v}\sin(vL)\right]\left(1-\dfrac{v^2}{\upkappa^2}\right)^{-1}h(v),       \\
A_{n/2}(v) = &\  e^{-i\rho}\frac{\upkappa}{4v}\left[e^{-ivL}\left(1+\frac{v}{\upkappa}\right)^{-n/2}\right.                              \\
             &\  \left.                            -e^{ ivL}\left(1-\frac{v}{\upkappa}\right)^{-n/2}\right]h(v)
\quad 
(n=3,5),                                                                                                                                 \\
A_2(v)     = &\  e^{-i\upkappa L}\left[\cos(vL)+i\frac{\upkappa}{v}\left(1+\frac{v^2}{\upkappa^2}\right)\sin(vL)\right]                  \\
             &\ \times \left(1-\dfrac{v^2}{\upkappa^2}\right)^{-2}h(v).
\end{align*}
The terms $j_{3/2}$ and $j_{5/2}$ are in fact nulls, since they are linear combinations of the vanishing integrals
\begin{eqnarray*}
\int_0^{2\pi}d\varphi\left.\partial_{\theta}\varPhi(\upkappa,\varphi,\theta)\right|_{\theta=0},
\quad
\int_0^{2\pi}d\varphi\left.\partial_{\theta}\varPhi(\upkappa,\varphi,\theta)\right|_{\theta=\pi}, \\
\int_0^{2\pi}d\varphi\left.\partial_{\theta}\varPhi_0    (\upkappa,\varphi,\theta)\right|_{\theta=0},
\quad
\int_0^{2\pi}d\varphi\left.\partial_{\theta}\varPhi_{\pi}(\upkappa,\varphi,\theta)\right|_{\theta=0}.
\end{eqnarray*}
The terms containing the factors proportional to $\cos(vL)$ and $v\sin(vL)$ decrease faster
than any power of $1/L$, according to the following classical Lemma (proved, e.g., in Ref.~\cite[p.~95]{Fedoryuk:1977}):

{\lemma[\rm localization principle]
\label{LocalizationPrinciple}
Let $S(x)\in C^{\infty}(\mathbb{R})$, $f(x)\in C^\infty_0(\mathbb{R})$ and $S'(x)\ne0$, $\forall x\in\mathbb{R}$. Then
\begin{equation*}
\int_{-\infty}^{\infty} dx f(x)\exp\left[i\lambda S(x) \right]=\mathcal{O}\left(\lambda^{-\infty}\right)
\enskip\text{as}\enskip \lambda\to+\infty.
\end{equation*}
}
Considering now that
\begin{equation*}
\lim_{L\to\infty}\frac{\sin(vL)}{{\pi}v}=\delta(v)
\end{equation*}
and the finite $L$ correction to the $\delta$ function is proportional to the integral
\begin{equation*}
\int_{-\infty}^\infty dv \left[h(v)\left(1-\dfrac{v^2}{\upkappa^2}\right)^{-1}-1\right]\frac{\sin(vL)}{v}
\end{equation*}
(which, according to Lemma~\ref{LocalizationPrinciple}, decreases faster than any power of $1/L$),
we can write:
\begin{align*}
J_{23}(\vec{L},\upkappa)
= &\  \frac{1}{8{\pi}L}\left[e^{-i{\upkappa}L}\varPhi(\upkappa,\varphi,0)+e^{i{\upkappa}L}\varPhi(\upkappa,\varphi,\pi)\right]  \\
  &\ +\frac{i\upkappa}{(4\pi\upkappa L)^2}\int_0^{2\pi}d\varphi\left[e^{-i{\upkappa}L}\varPhi_0    (\upkappa,\varphi,0) \right. \\
  &\  \left.                                                        +e^{ i{\upkappa}L}\varPhi_{\pi}(\upkappa,\varphi,0)\right]
     +\mathcal{O}\left(\frac{1}{L^3}\right).
\end{align*}
So that to obtain $J(\vec{L},\upkappa)$ we only have to sum up the terms $J_1(\vec{L},\upkappa)$ and $J_{23}(\vec{L},\upkappa)$; the result is
\begin{align}
\label{ResultOfSteps_4-5}
J(\vec{L},\upkappa)
= &\  e^{i{\upkappa}L}\left[\frac{\varPhi(-\upkappa\vec{l})}{4{\pi}L}
     +\frac{i}{8(\pi{\upkappa}L)^2}\int_0^{2\pi}d\varphi\varPhi_{\pi}(\upkappa,\varphi,0) \right. \nonumber \\
  &\ \left.+\mathcal{O}\left(\frac{1}{L^3}\right)\right].
\end{align}
Taking into account the definition of the function $\varPhi_{\pi}$ we get
\begin{align*}
\int_0^{2\pi} & d\varphi\varPhi_{\pi}(\upkappa,\varphi,0) \\
= &\ -\pi\left.\left(\frac{\partial^2}{\partial n_x^2}
     +\frac{\partial^2}{\partial n_y^2}+2\frac{\partial}{\partial n_z}\right)\varPhi\left(\upkappa\vec{n}\right)\right|_{\theta=\pi} \\
= &\ -{\pi}\upkappa^2\left.\left(\frac{\partial^2}{\partial q_x^2}
     +\frac{\partial^2}{\partial q_y^2}+\frac{2}{|\vec{q}|}\frac{\partial}{\partial q_z}\right)\varPhi\left(\vec{q}\right)\right|_{\vec{q}=-\upkappa\vec{l}}.
\end{align*}
Now, by combining the last equality with Eq.~\eqref{ResultOfSteps_4-5}, performing the back-rotation to the original coordinate system
and having regard to the rotation invariance of $\varPhi$, we arrive at the statement of the theorem.

\section{The higher-order corrections}
\label{sect:LessRigorousApproach}

The necessary condition for validity of the asymptotic formula \eqref{ExtendedGStheorem-Part-I} is obvious but,
as it was already mentioned in section \ref{sect:Introduction}, theorem~\ref{ExtendedGStheorem} does not yet provide
the necessary condition for the ISL breakdown, since the relative $\mathcal{O}(L^{-1})$ correction is imaginary if 
the function $\varPhi(\vec{q})$ is real-valued, as is actually the case in the models dealing with quasistable
external wave-packets. For example, in the above-mentioned CRGP model of Ref.~\cite{Naumov:2010um}
\begin{equation}
\label{Phi_CRGP}
\varPhi(\vec{q})  = \frac{\exp\left[-\dfrac{1}{4}\left(\widetilde{\Re}_s^{\mu\nu}\Delta^-_\mu\Delta^-_\nu 
                           +\widetilde{\Re}_d^{\mu\nu}\Delta^+_\mu\Delta^+_\nu\right)\right]}{(4\pi)^4\sqrt{|\Re_s||\Re_d|}},
\end{equation}
where $\Delta^\mp = q \mp q_{s,d}$ and the so-called inverse overlap tensors $\widetilde{\Re}_{s,d}$ (which define the effective space-time
overlap volumes of the external wave packets $\varkappa$) are expressed in terms of the most probable 4-velocities $u_\varkappa=p_\varkappa/m_\varkappa$
and momentum spreads $\sigma_{\varkappa}$ of the packets:
\begin{equation}
\label{InverseOverlapTensors}
\widetilde{\Re}_{s,d}^{\mu\nu} = \Big(\left|\left|\Re_{s,d}^{\mu\nu}\right|\right|^{-1}\Big)_{\mu\nu},
\quad
\Re_{s,d}^{\mu\nu}             = \sum_{\varkappa\in\mathcal{S},\mathcal{D}}\sigma_{\varkappa}^2\left(u_{\varkappa}^{\mu}u_{\varkappa}^{\nu}-g^{\mu\nu}\right).
\end{equation}
The summation in the last formula is over the sets $\mathcal{S}$ and $\mathcal{D}$ of the in and out wave packets in the source and detector vertices
of the macrodiagram describing the process. The explicit formulas for the overlap tensors are given in Appendix~1.
Substituting Eq.~\eqref{Phi_CRGP} into Eq.~\eqref{ExtendedGStheorem-Part-I} and neglecting the 4-momentum dependence of the $W$-boson propagators
(exact in the 4-fermion approximation) as well as the terms of the order of $1/({\upkappa}L)$ (small under any circumstance, see Eq.~\eqref{TheGScondition})
it can be shown that the leading correction term in Eq.~\eqref{ExtendedGStheorem-Part-I} is proportional to $i\upkappa/(\sigma_{\text{eff}}L)^2$, where
\begin{equation}
\label{InverseSquaredEffectiveMomentum-1}
\sigma_{\text{eff}}^{-2} = \frac{1}{4}\sum\limits_{i,j=1}^3\left(\widetilde{\Re}_s^{ij}+\widetilde{\Re}_d^{ij}\right)\left(\delta_{ij}-l_il_j\right).
\end{equation}
In particular, in the coordinate frame in which $\vec{l}=(0,0,1)$, the inverse squared effective momentum spread \eqref{InverseSquaredEffectiveMomentum-1}
is given by the relation
\begin{equation}
\label{InverseSquaredEffectiveMomentum-2}
\sigma_{\text{eff}}^{-2} = \frac{1}{4}\left(\widetilde{\Re}_s^{11}+\widetilde{\Re}_s^{22}
                                           +\widetilde{\Re}_d^{22}+\widetilde{\Re}_d^{22}\right),
\end{equation}
which clarifies the dependence of the spatial scale $\mathfrak{L}_0$ defined by Eq.~\eqref{L0} on the  4-velocities and momentum spreads
of the external wave packets. It is in particular seen that $\mathfrak{L}_0$ depends only on the transverse (with respect to the neutrino
propagation direction $\vec{l}$) components of the inverse overlap tensors.

In the general case, with a real-valued function $\varPhi(\vec{q})$, the function $|J(\vec{L},\upkappa)|^2$ still behaves
as $L^{-2}\left[1+\mathcal{O}\left(L^{-2}\right)\right]$ and in order to estimate the $\mathcal{O}(L^{-2})$ term,
one needs to calculate at least the second-order correction to the GS asymptotics.
In this section, we consider a development of the method used in Ref.~\cite{Korenblit:2013tya}.
The method allows us to derive the asymptotic expansion for the integral \eqref{GSintegral_Origin}
(as $L\to\infty)$, whilst under more restrictive conditions on the function $\varPhi(\vec{q})$
than those specified in theorem~\ref{ExtendedGStheorem}.

Namely, we will suppose that $\varPhi(\vec{q}) \in S(\mathbb{R}^3)$. 
Then $\varPhi(\vec{q})$ can be expressed in terms of the Fourier integral
\begin{equation}
\label{FourierTransformation}
\varPhi(\vec{q})= \int d\vec{x}e^{i\vec{q}\vec{x}}\widetilde{\varPhi}(\vec{x})
\end{equation}
and hence
\begin{equation}
\label{AfterFourierTransformation}
J(\vec{L},\upkappa)= \int d\vec{x}\widetilde{\varPhi}(\vec{x})I(\vec{x}-\vec{L},\upkappa),
\end{equation}
where
\begin{align*}
I(\vec{z},\upkappa)
& =  \int\frac{d\vec{q}}{(2\pi)^3}\frac{e^{i\vec{q}\vec{z}}}{\vec{q}^2-\upkappa^2-i\varepsilon} \\
& =  \frac{1}{  (2\pi)^2}\int_0              ^{\infty}\frac{dqq^2}{q^2-\upkappa^2-i\varepsilon}
     \int_0^{\pi}d\theta\sin\theta e^{iqz\cos\theta} \\
& =  \frac{1}{iz(2\pi)^2}\int_{-\infty}^{\infty}\frac{dq qe^{iqz}}{q^2-\upkappa^2-i\varepsilon}
  =  \frac{e^{i{\upkappa}z}}{4{\pi}z}.
\end{align*}
The order of integration in Eq.~\eqref{AfterFourierTransformation} was interchanged and the last equality was obtained
by applying the residue theorem. Thus
\begin{equation}
\label{Int_with_big_denominator}
J(\vec{L},\upkappa)
= \int\frac{d\vec{x}}{4\pi}\frac{e^{i\upkappa|\vec{L}-\vec{x}|}\widetilde{\varPhi}(\vec{x})}{|\vec{L}-\vec{x}|}.
\end{equation}
By expanding the function $e^{i\upkappa|\vec{L}-\vec{x}|}/|\vec{L}-\vec{x}|$ in Eq.~\eqref{Int_with_big_denominator}
in powers of $1/L$ (or, equivalently, in powers of the scalar variables $(\vec{l}\vec{x})$ and $(\vec{l}\times\vec{x})^2$)
we obtain after cumbersome but routine calculations:
\begin{equation}
\label{J_series}
J(\vec{L},\upkappa) = \frac{e^{i\upkappa L}}{4{\pi}L}\int d\vec{x}\widetilde{\varPhi}(\vec{x})e^{-i\upkappa(\vec{l}\vec{x})}
\sum_{k=0}^{\infty}\sum_{n=0}^{k}\sum_{m=0}^{\left[\frac{k-n}{2}\right]}\frac{\mathcal{D}_{knm}(\vec{x})}{L^k}\frac{\upkappa^n}{4^n},
\end{equation}
where
\begin{equation}
\label{D_knm}
\mathcal{D}_{knm}(\vec{x}) = (-1)^m i^nc_{knm}(\vec{l}\times\vec{x})^{2(n+m)}(\vec{l}\vec{x})^{k-n-2m}.
\end{equation}
The coefficients $c_{knm}$ in Eq.~\eqref{D_knm} are \emph{positive} numbers.
It is difficult to develop a general expression for these coefficients but it is an easy matter to derive $c_{knm}$
for moderately large values of $k$ sufficient for all practical applications. In particular, for $k\le6$ the coefficients are
\begin{align*}
c_{000} = &\ 1;            \enskip
c_{100} =    1,            \enskip
c_{110} =    2;            \\
c_{200} = &\ 1,            \enskip
c_{201} =    \frac{1}{2},  \enskip
c_{210} =    4,            \enskip
c_{220} =    2;            \\
c_{300} = &\ 1,            \enskip
c_{301} =    \frac{3}{2},  \enskip
c_{310} =    6,            \enskip
c_{311} =    \frac{3}{2},  \\
c_{320} = &\ 6,            \enskip
c_{330} =    \frac{4}{3};
\end{align*}
\begin{align*}
c_{400} = &\ 1,            \enskip
c_{401} =    3,            \enskip
c_{402} =    \frac{3}{8},  \enskip
c_{410} =    8,            \enskip
c_{411} =    6,            \\
c_{420} = &\ 12,           \enskip
c_{421} =    2,            \enskip
c_{430} =    \frac{16}{3}, \enskip
c_{440} =    \frac{2}{3}; 
\end{align*}
\begin{align*}
c_{500} = &\ 1,            \enskip
c_{501} =    5,            \enskip
c_{502} =    \frac{15}{8}, \enskip
c_{510} =    10,           \enskip
c_{511} =    15,           \\
c_{512} = &\ \frac{5}{4},  \enskip
c_{520} =    20,           \enskip
c_{521} =    10,           \enskip
c_{530} =    \frac{40}{3}, \enskip
c_{531} =    \frac{5}{3},  \\
c_{540} = &\ \frac{10}{3}, \enskip
c_{550} =    \frac{4}{15};
\end{align*}
\begin{align*}
c_{600} = &\ 1,            \enskip
c_{601} =    \frac{15}{2}, \enskip
c_{602} =    \frac{45}{8}, \enskip
c_{603} =    \frac{5}{16}, \enskip
c_{610} =    12,           \\
c_{611} = &\ 30,           \enskip
c_{612} =    \frac{15}{2}, \enskip
c_{620} =    30,           \enskip
c_{621} =    30,           \enskip
c_{622} =    \frac{15}{8}, \\
c_{630} = &\ \frac{80}{3}, \enskip
c_{631} =    10,           \enskip
c_{640} =    10,           \enskip
c_{641} =    1,            \enskip
c_{650} =    \frac{8}{5},  \\
c_{660} = &\ \frac{4}{45}.
\end{align*}
Finally, by applying the identity
\begin{equation*}
\bnabla_{\vec{q}}\varPhi(\vec{q})=i\int d\vec{x}\vec{x}e^{i\vec{qx}}\widetilde{\varPhi}(\vec{x})
\end{equation*}
which follows from Eq.~\eqref{FourierTransformation}, and interchanging the integration and summation%
\footnote{We do not discuss the legitimacy of this interchange.}
in Eq.~\eqref{J_series}, we arrive at the final asymptotic expansion for the integral \eqref{GSintegral_Origin}:
\begin{equation}
\label{final_formula}
J(\vec{L},\upkappa)=\frac{e^{i{\upkappa}L}}{4{\pi}L}
\sum_{k=0}^{\infty}\sum_{n=0}^{k}\sum_{m=0}^{\left[\frac{k-n}{2}\right]}
\frac{(-i)^k}{L^k}\left(\frac{\upkappa}{4}\right)^n\left[D_{knm}\varPhi(\vec{q})\right]_{\vec{q}=-\upkappa \vec{l}},
\end{equation}
where
\begin{align}
\label{final_formula_D}
D_{knm} = &\ \mathcal{D}_{knm}\left(-i\bnabla_{\vec{q}}\right) \nonumber \\
        = &\ (-1)^mc_{knm}\left(\vec{l}\times\!\bnabla_{\vec{q}}\right)^{2(n+m)}
                         \left(\vec{l}\bnabla_{\vec{q}}\right)^{k-n-2m}. 
\end{align}
Assuming that the function $\varPhi(\vec{q})$ is real-valued, the necessary conditions when the remainder after $k=2k'$ terms
in Eq.~\eqref{final_formula} can be neglected are
\begin{equation*}
\label{TheNecessaryConditions}
L^2 \gg \left|\frac{A_{k+2-\varsigma}}{A_{k-\varsigma}}\right|,
\quad
A_k = \sum_{n=0}^{k}\sum_{m=0}^{\left[\frac{k-n}{2}\right]}
\left(\frac{\upkappa}{4}\right)^n\left[D_{knm}\varPhi(\vec{q})\right]_{\vec{q}=-\upkappa \vec{l}},
\end{equation*}
where $\varsigma = 0,1$. It is seen that the leading order ($\propto L^{-2}$) correction in Eq.~\eqref{final_formula} coincides
with that in Eq.~\eqref{ExtendedGStheorem-Part-I} and (considering the identity 
$\left(\vec{l}\times\!\bnabla_{\vec{q}}\right)^2
=\bnabla_{\vec{q}}^2-\left(\vec{l}\bnabla_{\vec{q}}\right)^2$)
with that obtained in Ref.~\cite{Korenblit:2013tya}.

As we have mentioned in section~\ref{sect:Introduction}, in realistic models for the external wave packets with small momentum
spreads $\sigma_\varkappa$, the (real-valued) function $\mathcal{F}(q)$ appearing in Eq.~\eqref{4dGreenFunction} behaves like
a ``smeared'' 8-fold $\delta$-function (times a smooth factor) converging to $\text{const}\cdot\delta(q-q_s)\delta(q+q_d)$ as
$\sigma_\varkappa\to0$, $\forall\varkappa$; an example is given by Eq.~\eqref{Phi_CRGP}.
Let us now limit ourselves to the case of light neutrinos with the masses $m=m_i$ ($i=1,2,\ldots$) satisfying the
ultrarelativistic conditions
\[
(q_s^0)^2 \approx (q_d^0)^2 \gg m_i^2.
\] 
After substituting Eqs.~\eqref{final_formula} and \eqref{final_formula_D} into Eq.~\eqref{4dGreenFunction}, the remaining integration
in variable $q_0$ can be performed by the saddle-point method and the corresponding stationary point of the integrand can be expressed
as a Taylor series in powers of $(m_i^2/q_s^0)^2$ with the leading-order term approximately equal to $q_s^0$.
As a result, the functions
\[
\mathfrak{J}(\vec{L},\upkappa) = 4{\pi}e^{-i{\upkappa}L}J(-\vec{L},\upkappa)
\]
(with $\left.\varPhi(\vec{q})\equiv\mathcal{F}(q)\right|_{\vec{q}=-\upkappa\vec{l}}$)
for each neutrino mass are essentially equal to each other and, moreover, they can be pulled out of the integral \eqref{4dGreenFunction}
in the point $\upkappa = q_s^0 \approx q_d^0$.
Therefore the integral \eqref{4dGreenFunction} is to a good precision proportional to the universal (neutrino mass independent) function
$\mathfrak{J}(\vec{L},q_s^0)$ and the squared absolute value of the full amplitude is thus proportional to $|\mathfrak{J}(\vec{L},q_s^0)|^2$.
Note that, in this approximation, the corrections to the GS asymptotics do not change the neutrino oscillation pattern, including
the field-theoretical decoherence effects which become essential at large distances.
It is easy to prove that
\begin{align}
\label{ProbabilitySeries}
\left|\mathfrak{J}(\vec{L},\upkappa)\right|^2 
= &\ \sum_{k=1}^\infty\sum_{n=0}^{2(k-1)}\frac{C_{kn}\upkappa^n}{L^{2k}} \nonumber \\
= &\ \frac{1}{L^2}\left[\varPhi^2(\upkappa\vec{l})+
  \sum_{k=1}^\infty\sum_{n=0}^{2k}\frac{C_{k+1,n}\upkappa^n}{L^{2k}}\right].
\end{align}
The series in the right-hand part of Eq.~\eqref{ProbabilitySeries} describes the deviation from the classical ISL
in the neutrino-induced event rate at short distances satisfying however the necessary conditions
\[
L^2 \gg \left|\sum_{n=0}^{2k}C_{k+1,n}\upkappa^n\!\Big/\sum_{n=0}^{2(k-1)}C_{kn}\upkappa^n\right|
\]
which restrict the range of validity of the truncated asymptotic series.
Using the coefficients $c_{knm}$ derived above for $k\le6$ allow us to calculate the low-order coefficient functions
$C_{kn}=C_{kn}(\upkappa,\vec{l})$ for $k\le3$. We obtain
\begin{align*}
C_{10} = &\  f^2; \\
C_{20} = &\  ff_{20}-2ff_{02}+f_{01}^2,
\enskip 
C_{21} =     f_{20}f_{01}-2ff_{21}, \\
C_{22} = &\    \frac{1}{4}\left(f_{20}^2-ff_{40}\right);                                                                                                   \\
C_{30} = &\  \left(\frac{3}{4}f_{40}-6f_{22}+2f_{04}\right)f+\left(3f_{21}-2f_{03}\right)f_{01} \\
         &\ +\left(f_{02}-\frac{1}{2}f_{20}\right)^2,                    \\
C_{31} = &\   \left(4f_{23}-3f_{41}\right)f+\left(\frac{1}{2}f_{20}+2f_{02}\right)f_{21} \\
         &\ +3\left(\frac{1}{4}f_{40}-f_{22}\right)f_{01}-f_{20}f_{03},          \\
C_{32} = &\   \frac{1}{4}\left(f_{40}-6f_{22}\right)f_{20}+\frac{1}{4}f_{02}f_{40}+\frac{1}{4}\left(6f_{42}-f_{60}\right)f \\
         &\  -\frac{3}{4}f_{01}f_{41}+f_{21}^2, \\
C_{33} = &\  -\frac{1}{24}\left(9f_{41}f_{20}+f_{60}f_{01}-4ff_{61}-6f_{40}f_{21}\right),                                                                   \\
C_{34} = &\ \frac{1}{192}\left(ff_{80}+3f_{40}^2-4f_{20}f_{60}\right);
\end{align*}
where
\begin{equation*}
f_{ij} = \left.\left(\vec{l}\times\!\bnabla_{\vec{q}}\right)^i\left(-\vec{l}\bnabla_{\vec{q}}\right)^j
         \varPhi(\vec{q})\right|_{\vec{q}=\upkappa\vec{l}},
\quad
f \equiv f_{00}=\varPhi(\upkappa\vec{l}).
\end{equation*}

\section{The ISL violation}
\label{sect:TheISLviolation}

By using the CRGP model of Ref.~\cite{Naumov:2010um} and taking into account Eq.~\eqref{ProbabilitySeries}, one can derive
the formula for the event rate in the detector. The formula for the rate has exactly the same form as Eq.~\eqref{EventRate}
with the only major difference that now
\begin{eqnarray}
\label{TheISLviolatinFactor}
d\mathfrak{F}_\nu \propto \frac{1}{|\vec{y}-\vec{x}|^2}
\Bigg[1+{\sum_{n\ge1}\frac{\mathfrak{C}_n}{|\vec{y}-\vec{x}|^{2n}}}\Bigg],
\end{eqnarray}
where the coefficients $\mathfrak{C}_n$ are defined as
\begin{align*}
\label{CorrectionFactorToProbabilityCoefficients}
\mathfrak{C}_n = &\ 2(-1)^{n}\left[\sum_{k=1}^{n}(-1)^k\mathscr{C}_k\mathscr{C}_{2n-k}+\mathscr{C}_{2n}\right]-\mathscr{C}_n^2, \\
\mathscr{C}_k  = &\ \sum_{n=0}^{k}\sum_{m=0}^{\left[\frac{k-n}{2}\right]}
 \left(\frac{\upkappa}{4}\right)^n\left[\frac{D_{knm}\varPhi(\vec{q})}{\varPhi(\vec{q})}\right]_{\vec{q}=E_\nu\vec{l}}.
\end{align*}
The series in the right-hand part of Eq.~\eqref{TheISLviolatinFactor} obviously leads to a \emph{deviation from the inverse-square law}.

It is worth to note that such deviation can be physically interpreted in a different and more transparent way,
by exploiting the concept of a duality between the propagator and wave-packet descriptions of the neutrino production
and detection process. Namely, the corrections to the GS asymptotics can be treated as a change of probabilities
of neutrino production and detection. This interpretation is briefly described in Appendix~2.

In the domain of applicability of the asymptotic expansion \eqref{ProbabilitySeries}, the sign of this deviation is governed
by the sign of the leading coefficient 
\begin{align*}
\mathfrak{C}_1
= &\  \mathscr{C}_1^2-2\mathscr{C}_2 \\
= &\  \frac{1}{f^2}\Big[f_{01}^2+f\left(f_{20}-2f_{02}\right) 
     +\left(f_{20}f_{01}-2ff_{21}\right)E_\nu \\
  &\ +\frac{1}{4}\left(f_{20}^2-ff_{40}\right)E_\nu^2\Big].
\end{align*}
It is not easy to determine this sign in the most general case.
The physical range of applicability of the series \eqref{final_formula}, \eqref{ProbabilitySeries},
and \eqref{TheISLviolatinFactor} crucially depends on the model of the external wave packets which defines the explicit form of
the function $\mathcal{F}(q)$ and thereby the scale of the ISL breakdown.
Moreover, in a real environment, the potentially observable effects are dependent on the reactions (particle content and dynamics)
in the source and detector vertices, on the statistical distributions of the in-packets over configuration space, phase space,
polarizations, etc., and on many other experimental circumstances. Without specification of all these details,
one can do no more than speculate on feasible manifestation of the ISL breakdown.
Nevertheless, from a detailed numerical analysis of the simplest particular subprocesses $1\to2$, $1\to3$ and $2\to2$
(for which purpose the general formulas, obtained within the CRGP model were used, see Appendix~1) we observed that
$\mathfrak{C}_1<0$ either everywhere or at least in the essential region of the physical phase space of these subprocesses.
Under the assumption that this is the case for the real-world experimental environment,
\emph{in the range of applicability of the asymptotic series \eqref{TheISLviolatinFactor}} the ISL violation leads
to a \emph{decrease} in the neutrino event rate \eqref{EventRate}.
Our numerical analysis shows that the scope of ISL violation is actually defined by the scale parameter \eqref{L0},
which varies within very wide limits and can become macroscopically large for the appropriate combination of
the neutrino energy and momentum spreads $\sigma_\varkappa$ of the external wave packets.
Since the real value of the ISL corrections drastically depends on the momentum distributions of the external particles
and dynamics of their interactions, this value cannot be predicted without a knowledge of this environment.

Considerable recent attention has been focused on the hypothesis of light sterile neutrinos with eV-range masses.
One of the reasons for the interest in this hypothesis is a statistically significant deficit of the count rates
of electron neutrinos from intense radioactive sources used for calibration of the Ga-Ge solar neutrino detectors
GALLEX \cite{Anselmann:1994ar,Hampel:1997fc,Kaether:2010ag} and SAGE \cite{Abdurashitov:1996dp,Abdurashitov:1998ne,Abdurashitov:2005tb}
(with the typical $L$ of about 0.5--2~m) 
and a deficit (at more than two standard deviation significance) in the measured capture rate of electron antineutrinos
from nuclear fission, revealed after a recent re-analysis of the past SBL ($8~\text{m} \lesssim L \lesssim 100~\text{m}$)
reactor experiments \cite{Mention:2011rk,Mueller:2011nm,Huber:2011wv,Lasserre:2012jc,Lasserre:2011zz,Mention:2013kra,Zhang:2013ela}.
Both these anomalies are compatible with the mixing between the standard active and eV-scale sterile neutrinos that allow
some of the electron (anti)neutrinos from the source to ``disappear'' avoiding their interaction with the detector
\cite{Conrad:2012qt,Giunti:2012tn,Kang:2013gpa,Kopp:2013vaa,Giunti:2013rfa}.
There is and yet a number of results coming from accelerator experiments which disfavor this interpretation
at least within the simplest ``3+1'' four-neutrino scheme (for the recent reviews, see, e.g., Refs.~\cite{Abazajian:2012ys,Kopp:2013vaa}).
Moreover, the additional light neutrino species seem to be in some conflict with the minimal $\Lambda$CDM
cosmological model \cite{Hamann:2011ge,Abazajian:2012ys,Mirizzi:2013kva}
and potentially with the recent high-precision data from the medium baseline reactor experiments Daya Bay \cite{An:2012eh,An:2012bu}
and RENO \cite{Ahn:2012nd} operating at the distances longer than about $400$~m
(see, however, the recent analyses \cite{Palazzo:2013bsa,Giunti:2013aea,Esmaili:2013yea} and references therein). 
The described controversial situation suggests to investigate alternative (or complimentary) explanations of the
intriguing anomalies in the gallium and reactor experiments.

The ISL violation provided by high-order corrections to the GS asymptotics seems to be an attractive alternative explanation
of the one or even both of these anomalies, since it does not require any ``new physics'' and can, with reasonable facility,
be tested experimentally.
In particular, the SBL experiments (both completed and currently underway) to search for the light sterile neutrinos with
the reactor and accelerator (anti)neutrino beams, including the experiments with intense $\beta$-active sources
(see, e.g., Refs.~\cite{Furuta:2011iu,Dwyer:2011xs,Spitz:2012gp,Cucoanes:2012jv,Derbin:2012kf,Serebrov:2012sq,Belov:2013qwa,Elnimr:2013}
and also Refs.~\cite{Yasuda:2011np,Gaffiot:2013bba} for reviews and further references)
are trivially adaptable to test the ISL violation effects.
The expected signatures of the sterile neutrino and ISL violation are quite different. First of all, the ISL violation effect
does not affect the measured event rate at the distances $L\gg\mathfrak{L}_0$, while the sterile neutrinos with the eV-scale masses
lead to a decrease in the event rate at any distance longer than the corresponding oscillation length and the \emph{relative} effect
(defined as the ratio of the averaged survival probabilities of electron (anti)neutrino in the $3\nu$ and $4\nu$ mixing schemes)
becomes nearly independent of $L$ at distances above $50-100$~m (see, e.g., Ref.~\cite{Kopp:2013vaa}).
Next, in the not too short baseline region, where the asymptotic series \eqref{TheISLviolatinFactor} is presumably approximated
by the leading term, the ISL violation effect is given by the very simple factor
\[
1-L_0^2/L^2,
\]
where $L_0\sim\langle\mathfrak{L}_0\rangle$ is a neutrino energy dependent parameter of dimension of length.
This behavior is quite distinguishable from the oscillation pattern expected in the typical active-sterile neutrino mixing scenarios.
But of course the signature of the ISL violation will be more intricate if it takes place together with the sterile neutrino contribution.

By using the available SBL reactor data it is easy to estimate that if, indeed, the ISL violation is responsible for the reactor anomaly
either totally or partially (i.e.\ in parallel with the light sterile neutrinos) the parameter $L_0$ should be of the order of $2-4$~m,
which (very roughly) corresponds to the effective momentum spread, $\sigma_{\text{eff}}$, of the order of $0.1-10$~eV.%
\footnote{A more definitive estimate is not feasible for the moment, considering that the function $\sigma_{\text{eff}}$
          varies within wide limits over the phase spaces of the nuclear fission ($\overline{\nu}_e$-produced)
          processes in the power reactor and the inverse $\beta$-decay reaction in the detector.}
If this is the case, then the ISL violation is essentially unobservable at $L \gtrsim 100$~m and thus it does not at all
interfere with the usual three-flavour oscillations.
If however it is relevant to the gallium (rather than reactor) anomaly, the parameter $L_0$ should be much smaller.
At the same time it must be emphasized that the gallium anomaly generally requires more complicated analysis, considering that
in the typical calibration experiments the ``source'' and ``detector'' dimensions are compatible with each other as well as
with the distance between them and thus the simplified formula \eqref{EventRate} is definitely too rough approximation.
Moreover, the asymptotic series \eqref{TheISLviolatinFactor} can be, and most probably is, inapplicable to so short distances.
Any case, best suited experiments for investigating the possible ISL violation effect would be the experiments
with variable short or very short baseline, holding the latter to be nonetheless sufficiently large in comparison with
the dimensions of the source and detector.

\section{Conclusions}
\label{sect:Conclusions}

In this work the extended GS theorem is proved and the next-to-leading correction to the GS asymptotics are found
in terms of an asymptotic expansion of the generalized neutrino propagator (and thus the neutrino flavor transition amplitude)
in powers of $1/L$.
It is argued that within the domain of applicability of the asymptotic series~\eqref{final_formula} and \eqref{ProbabilitySeries},
the power corrections do not affect the flavor transition probability but can lead to a measurable violation of the classical
inverse-square law for the neutrino flux.
The possibility is discussed that this violation has already appeared in the past reactor or Ga-Ge (anti)neutrino experiments.
This hypothesis can be tested in the dedicated short or very short baseline experiments with nuclear fission reactors
or radioactive neutrino sources.

\begin{acknowledgements}
This work was supported by the Federal Target Program ``Scientific and Scientific-Pedagogical Personnel of the Innovative Russia''
under Contract No.~2012-1.5-12-000-1011-008.
The authors would like to thank Z.~G. Berezhiani, S.~E. Korenblit, and especially D.~V. Naumov (who was the active partner
in the early stages of this research) for very useful suggestions, comments and criticisms.
\end{acknowledgements}

\section*{Appendix 1. Inverse overlap tensors.}
\label{Appendix-1}

The higher order corrections are rather involved but, within the CRGP model for the external wave packets, they are constructed from the spatial
components of inverse overlap tensors $\widetilde{\Re}_s^{\mu\nu}$ and $\widetilde{\Re}_d^{\mu\nu}$ defined by Eqs.~\eqref{InverseOverlapTensors}.
Here we write down the explicit expressions for these components. It is proved that for arbitrary lepton number violating processes
\begin{equation}
\label{TheProcess}
I_s \oplus I_d \to F_s \oplus F_d
\end{equation}
(where $I_s$ and $I_d$ denote the sets of incoming wave packets in the source and detector vertices, respectively; and, similarly, $F_s$ and $F_d$
denote the sets of outgoing wave packets which include charged leptons) the components of the inverse overlap tensors can be written as
\begin{equation*}
\widetilde{\Re}_{s,d}^{\mu\nu} 
=  \frac{1}{|\Re_{s,d}|}\sum\limits_{a,b,c\in\mathcal{S},\mathcal{D}}
   \sigma_{a}^2\sigma_{b}^2\sigma_{c}^2\mathfrak{I}^{abc\,\mu\nu}_{s,d},
\end{equation*}
where
\begin{align*}
\mathfrak{I}^{abc\,00}_{s,d}
= &\  \left[\varGamma_a\varGamma_b-\frac{1}{3}(u_au_b)\right](u_bu_c)(u_cu_a)                          \\
  &\ +\frac{1}{2}\varGamma_a^2\left[1-(u_bu_c)^2\right]+\frac{1}{3},                                   \\
\mathfrak{I}^{abc\,0i}_{s,d}
= &\  \frac{1}{2}\varGamma_c\left[(u_a u_b)^2u_{c i}-u_{c i}-2(u_a u_b)(u_b u_c)u_{a i}\right.         \\
  &\  \left. +2(u_a u_c)u_{a i}\right]-\varGamma_b(u_a u_b)u_{a i},                                    \\
\mathfrak{I}^{abc\,ij}_{s,d}
= &\  \varGamma_a\varGamma_b\left[(u_cu_a)(u_cu_b)-(u_au_b)\right]\delta_{ij}                          \\
  &\ +\frac{1}{2}\left(\varGamma_{c}^2-1\right)\left[1-(u_au_b)^2\right]\delta_{ij}                    \\
  &\ +\left\{\varGamma_{c}\left[\varGamma_{a}(u_b u_c) \right.                                              
     +\varGamma_{b}(u_c u_a)-\varGamma_{c}(u_a u_b)\right]- \varGamma_{a}\varGamma_{b}                 \\
  &\ \left. +(u_a u_b)\right\}u_{ai}u_{bj}                                                             
     +\varGamma_{b}\left[\varGamma_{b}-\varGamma_{a}(u_a u_b)\right]u_{ci}u_{cj},                      \\
|\Re_{s,d}|
= &\  \sum\limits_{a,b,c,d\in\mathcal{S},\mathcal{D}}\!\!\!\sigma_{a}^2\sigma_{b}^2\sigma_{c}^2\sigma_{d}^2
      \left\{\frac{1}{3}\left[(u_au_b)(u_bu_c)(u_cu_a)-1\right] \right.                                \\
  &\ -\frac{1}{2}\varGamma_a\varGamma_b(u_au_b)\left[(u_cu_d)^2-1\right]                               \\                   
  &\ +\varGamma_a\varGamma_b(u_cu_a)\left[(u_cu_d)(u_du_b)-(u_bu_c)\right]                             \\
  &\  \left.+\frac{1}{6}\varGamma_d^2\left[3(u_bu_c)^2-2(u_au_b)(u_bu_c)(u_cu_a)-1\right]\right\};
\end{align*}
the symbols $\mathcal{S}=I_s{\oplus}F_s$ and $\mathcal{D}=I_d{\oplus}F_d$ denote the sets of the in and out packets in the source and
detector vertices, respectively; $u_a=p_a/m_a=(\varGamma_a,\vec{u}_a)$ and $\varGamma_a$ are the (most probable) 4-velocity and Lorentz factor
of the packet $a$, respectively. It can  be verified that the determinant $|\Re_{s,d}|$ is invariant and $\widetilde{\Re}_{s,d}^{\mu\nu}$ transforms
as a tensor under Lorentz transformations. Note that the above formulas were derived without taking into account the energy-momentum conservation.

As the simplest illustration, we consider here the particular case: the two-particle decay $a\to\ell+\nu^*$ in the source,
where $\ell$ is a charged lepton and $\nu^*$ is a virtual (anti)neutrino with definite mass. In this case
\begin{align*}
\label{Two-part}
\widetilde{\Re}_{s}^{\mu\nu}
= &\  \Big\{\sigma_a^4u_{a}^{\mu}u_{a}^{\nu}+\sigma_\ell^4u_{\ell}^{\mu}u_{\ell}^{\nu} 
     -\sigma_a^2\sigma_\ell^2\left\{ g^{\mu\nu}\left[(u_au_\ell)^2-1\right]\right.            \\
  &\  \left.  -(u_au_\ell)(u_{a}^{\mu}u_{\ell}^{\nu}+u_{\ell}^{\mu}u_{a}^{\nu})\right\}\Big\} \\
  &\  \times  \left\{\sigma_a^2\sigma_\ell^2\left(\sigma_a^2+\sigma_\ell^2\right)\left[(u_au_\ell)^2-1\right]\right\}^{-1}.
\end{align*}
This simple expression can also be used for estimations of the inverse overlap tensors for more involved reactions and decays
in the case of strong hierarchy between the momentum spreads $\sigma_\varkappa$.
Let us mention, in passing, that the inequality $\mathfrak{C}_1<0$ (see Sect.~\ref{sect:TheISLviolation}) holds over the whole phase space
of the two-particle decay into real (on-shell) neutrino, assuming exact energy-momentum conservation.
The corresponding contribution to the inverse squared effective momentum spread \eqref{InverseSquaredEffectiveMomentum-2}
is then given by the following expression:
\begin{equation*}
\frac{1}{4}\left(\widetilde{\Re}_s^{11}+\widetilde{\Re}_s^{22}\right)
= \frac{K}{\left(2E_\nu^{\star}\right)^2}
  \left(\frac{m_a^2}{\sigma_a^2}+\frac{m_\ell^2}{\sigma_\ell^2}\right)+\frac{1}{2\left(\sigma_a^2+\sigma_\ell^2\right)},
\end{equation*}
in which
\begin{equation*}
K = \left(1-\frac{E^{-}_{\nu}}{E_{\nu}}\right)\left(\frac{E^{+}_{\nu}}{E_{\nu}}-1\right),
\end{equation*}
$E^{\pm}_{\nu} = E_\nu^{\star}\varGamma_a(1\pm|\mathbf{v}_a|)$ are the kinematic boundaries of $E_\nu$,
$E_\nu^{\star}=\left(m_a^2-m_\ell^2\right)/(2m_a)$ is the neutrino energy in the rest frame of the decaying
particle $a$, and $\mathbf{v}_a$ is the velocity of $a$. The neutrino mass has been neglected in derivation.
The kinematic factor $K$ varies within wide limits at relativistic energies but vanishes when the particle
$a$ is at rest. For the median neutrino energy, $E_\nu=\overline{E}_\nu = \left(E^{+}_{\nu}+E^{-}_{\nu}\right)/2=E_\nu^{\star}\varGamma_a$,
\[
K = |\mathbf{v}_a|^2 = 1-\left({E_\nu^{\star}}/{\overline{E}_\nu}\right)^2.
\]
\section*{Appendix 2: Effective neutrino wave packet}
\label{sec:NeutrinoWavePacket}

Here we consider some kind of dualism between the propagator and effective wave-packet
approaches. Namely we show that in the asymptotic regime $L\to\infty$, the generalized
neutrino Green function \eqref{4dGreenFunction} can be represented as a product of
outgoing and incoming effective wave functions of the \emph{on-mass-shell} neutrino.
Below, we use the results of Ref.~\cite{Naumov:2010um} (obtained within the CRGP model)
but with a different and -- as we believe -- more adequate interpretation which, in particular,
allows us to get another look at the corrections to the GS asymptotic.

Let us consider the factor
\begin{equation}
\label{NeutrinoWaveFunctionProduct}
\frac{1}{L}e^{-\varOmega(T,L)}P_-\left(\hat{p}_{\nu}+m\right)P_+
\end{equation}
originated in the full amplitude after applying the GS theorem \ref{TheGStheorem} to
Eq.~\eqref{4dGreenFunction} and using the saddle-point integration in variable $q_0$~\cite{Naumov:2010um}.
Here
\begin{equation}
\label{varOmega_inv}
\varOmega(T,L) = i(p_{\nu}X)+\frac{2\mathfrak{D}^2}{E_\nu^2}\left[(p_{\nu}X)^2-m^2X^2\right],
\end{equation}
$p_{\nu}= (E_{\nu},\vec{p}_{\nu})$ is the mean neutrino 4-momentum ($\vec{p}_{\nu}\approx E_{\nu}\vec{l}$
in the ultrarelativistic case) and $P_{\pm}=\tfrac{1}{2}(1\pm\gamma_5)$ are the chiral projectors involved
into the Standard Model amplitude. The 4-vector $X=(X_0,\vec{X})=X_d-X_s$ is the difference between the
so-called impact points $X_s$ and $X_d$ which determine the coordinates of the space-time overlap regions
of the external wave packets in the source and detector vertices; note that $\vec{X}=\vec{L}=L\vec{l}$
and $X_0=T$ (see Ref.~\cite{Naumov:2010um} for more details).
The function $\mathfrak{D}$, which can be treated as the uncertainty of the virtual neutrino energy, is defined by
\begin{align}
\label{Fussiness}
\frac{1}{\mathfrak{D}^2}
   =   &\ \left[\frac{d^2\ln\mathcal{F}(q)}{dq_0^2}\right]_{q=q_0^{\text{st}}} \nonumber \\
\simeq &\ 2\left(\widetilde{\Re}_s^{\mu\nu}+\widetilde{\Re}_d^{\mu\nu}\right)l_{\mu}l_{\nu}\left[1+\mathcal{O}\left(\frac{m^2}{E_\nu^2}\right)\right],
\end{align}
where $q_0^{\text{st}} \approx E_{\nu}$ is the stationary point.
So, in the first approximation the function $\mathfrak{D}$ does not depend on the neutrino mass and
in the coordinate frame where $\vec{l} = \left(0,0,1\right)$ is determined by the $00$, $03$ and $33$ components
of the inverse overlap tensors. By applying the identity
\[
P_-\left(\hat{p}_{\nu}+m\right)P_+=P_-u(\vec{p}_{\nu})\overline{u}(\vec{p}_{\nu})P_+,
\]
in which $u(\vec{p}_\nu)$ is the usual Dirac bispinor for a free ultrarelativistic
left-handed neutrino, we can rewrite Eq.~\eqref{NeutrinoWaveFunctionProduct} as follows
\begin{equation}
\label{Factorization}
u(\vec{p}_{\nu})\frac{e^{-\varOmega(T,L)}}{L} \overline{u}(\vec{p}_{\nu})
= \frac{\uppsi_{X_d}(\vec{p}_{\nu},X_s-X_d)\overline{\uppsi}_{X_s}(\vec{p}_{\nu},X_d-X_s)}{|\vec{X}_d-\vec{X}_s|},
\end{equation}
where
\begin{align*}
\uppsi_{y}(\vec{p}_{\nu},x)
  =    &\ \exp\left\{-i(p_{\nu}y)-\frac{\mathfrak{D}^2}{E_\nu^2}\left[(p_{\nu}x)^2-m^2x^2\right]\right\}u(\vec{p}_{\nu}) \\
\equiv &\ \Xi_{y}(\vec{p}_{\nu},x)u(\vec{p}_{\nu})
\end{align*}
and
\begin{equation*}
\overline{\uppsi}_{y}(\vec{p}_{\nu},x) 
= \uppsi^\dag_{y}(\vec{p}_{\nu},x)\gamma_0
= \overline{u}(\vec{p}_{\nu})\Xi_{y}^*(\vec{p}_{\nu},x).
\end{equation*}
Let us now compare the function $\uppsi_{y}(\vec{p}_{\nu},x)$ with the wave function
\begin{equation}
\label{GenericFermionWaveFunction}
\psi_y(\vec{p},s,x) = \langle{0}|\Psi(x)|\vec{p},s,y\rangle
\end{equation}
describing a generic fermion wave packet $|\vec{p},s,y\rangle$ in the coordinate representation
(here $\Psi(x)$ is the free fermion field and $s$ is the spin projection).
Within the CRGP approximation, valid under the conditions 
\[
\sigma^2\ll m^2, \quad \sigma^4(px)^2\ll m^4, \quad \sigma^4\left[(px)^2-m^2x^2\right] \ll m^4,
\]
the function \eqref{GenericFermionWaveFunction} is \cite{Naumov:2010um}
\begin{equation*}
\psi_y(\vec{p},s,x) \approx u_s(\vec{p})\exp\left\{-i(py)-\frac{\sigma^2}{m^2}\left[(px)^2-m^2x^2\right]\right\}.
\end{equation*}
We see therefore that the spinor multiplier $\uppsi_{X_d}(\vec{p}_{\nu},X_s-X_d)$ in Eq.~\eqref{Factorization}
can be interpreted as the wave function describing the wave packet of a \emph{real} (on-shell) neutrino incoming
to the detector, while the multiplier $\uppsi_{X_s}^\dag(\vec{p}_{\nu},X_d-X_s)/|\vec{X}_d-\vec{X}_s|$
can naturally be treated as the spherical wave function of the same neutrino packet escaping the source.
From this interpretation it in particular follows that the effective neutrino momentum spread, 
$\sigma_{\nu} = m\mathfrak{D}/E_{\nu}$, is not a constant but an invariant function of the
neutrino energy as well as of the mean momenta, masses, and momentum spreads of the external wave packets;
due to the factor $m/E_{\nu}$ it is extremely small in all realistic situations.
The main distinctive feature of the effective neutrino wave packet from the conventional quantum-mechanical
wave packets is that it depends on both the source and detector variables. Therefore the duality under discussion
should not be understood in a literal sense.

The case becomes even more complicated when one takes into account the higher-order corrections to the GS asymptotics.
These corrections yield an additional factor 
\begin{equation}
\label{SuppressionFactor}
\Sigma = 1+\sum_{k=1}^\infty\frac{a_k}{L^k}
\end{equation}
with the complex-valued coefficient functions $a_k$ dependent on the contributions from both the source and detector.
Except for very special or trivial cases (when, i.g., one of the contributions is negligible) the functions $a_k$
cannot be factorized to the product of the source and detector dependent multipliers.
Therefore, if the spacing between the impact points is short the incoming and outgoing neutrino wave packets
are not yet separated from each other as it is the case at the asymptotically large distances.
In other words, the duality between the propagator and wave-packet treatments is destroyed at short distances
and such a case cannot be adequately described in terms of the asymptotically free in and out neutrino wave packets.

Considering now that the modulus-squared full amplitude incorporates the factors $|\Xi_{X_s}|^2$ and $|\Xi_{X_d}|^2$
which represent (up to a normalization) the probability densities to obtain the in and out neutrino wave packets
near the corresponding impact points in the source and detector, we conclude that the factor $|\Sigma|^2$ affects 
\emph{both} these densities or, to put this another way, at relatively short distances between the impact
points the factor $|\Sigma|^2$ changes the probabilities of neutrino production and detection.

It is important to note that, in contrast with the function $\mathfrak{D}$, the coefficient functions $a_k$
include the components $\widetilde{\Re}_{s,d}^{11}$ and $\widetilde{\Re}_{s,d}^{22}$ transversal with respect
to the neutrino momentum direction $\vec{l}=(0,0,1)$ (see Eq.~\eqref{InverseSquaredEffectiveMomentum-2}
as the simplest example). These components are not suppressed by the Lorentz factor $E_{\nu}/m$ and thus
the ISL violation effect can be large even for massless neutrinos.

\bibliography{references}

\end{document}